\documentclass{article}
\usepackage[T1]{fontenc} % add special characters (e.g., umlaute)
\usepackage[utf8]{inputenc} % set utf-8 as default input encoding
\usepackage{ismir,amsmath,cite,url}
\usepackage{graphicx}
\usepackage{color}
\usepackage{booktabs}
\usepackage[multiple]{footmisc}
\usepackage{array,makecell}
\usepackage{flushend}
\usepackage{tabularx}
\usepackage{CJKutf8}
    \newcolumntype{L}{>{\raggedright\arraybackslash}X}

\title{PolySinger: Singing-Voice to Singing-Voice Translation from English to Japanese}

\oneauthor
  {Silas Antonisen, Iv\'an L\'opez-Espejo} {Department of Signal Theory, Telematics and Communications, University of Granada, Spain \\ {\tt \{santon,iloes\}@ugr.es}}

\begin{document}

\maketitle
\begin{abstract}
The speech domain prevails in the spotlight for several natural language processing (NLP) tasks while the singing domain remains less explored. The culmination of NLP is the speech-to-speech translation (S2ST) task, referring to translation and synthesis of human speech. A disparity between S2ST and the possible adaptation to the singing domain, which we describe as singing-voice to singing-voice translation (SV2SVT), is becoming prominent as the former is progressing ever faster, while the latter is at a standstill. Singing-voice synthesis systems are overcoming the barrier of multi-lingual synthesis, despite limited attention has been paid to multi-lingual songwriting and song translation. This paper endeavors to determine what is required for successful SV2SVT and proposes PolySinger (\textbf{Poly}glot \textbf{Singer}): the first system for SV2SVT, performing lyrics translation from English to Japanese. A cascaded approach is proposed to establish a framework with a high degree of control which can potentially diminish the disparity between SV2SVT and S2ST. The performance of PolySinger is evaluated by a mean opinion score test with native Japanese speakers. Results and in-depth discussions with test subjects suggest a solid foundation for SV2SVT, but several shortcomings must be overcome, which are discussed for the future of SV2SVT.
\end{abstract}
\section{Introduction}\label{sec:introduction}
Speech-to-speech translation (S2ST) is a method for translating human speech into another language using synthetic speech. To do this, the conventional approach is to concatenate technologies that process separate parts of human speech into a complete system, where the cornerstones are speech recognition, machine translation and speech synthesis \cite{janus-iii,ATR,ITU-T}. Although the use of end-to-end (E2E) solutions for S2ST has been studied thanks to the emergence of seq-to-seq models \cite{translatatron,textlessS2ST,seamlessm4t2023}, neither E2E nor cascaded solutions have been attempted in the singing domain.

Singing-voice synthesis (SVS) systems have in recent years become very capable of human-like singing \cite{liu2022diffsinger,nnsvs} and have even accomplished multi-lingual synthesis\cite{crosssinger}. However, while the synthetic voice can sing cross-lingually, the songwriter might not be able to write cross-lingually. 

%Even when provided foreign-language lyrics, the synthetic voice must be tuned by manual note manipulation instead of by a melody and rhythm delivered by a vocal performance in the performer's native language.

%Generally, lyrics translation is a more involved task, as the translator must assess how much of the meaning may be altered for the song to remain as easily singable as the original one. Many cultural expressions can also be apparent in a verse, which requires the translator to fathom extensive amounts of cultural knowledge in order to convey the same meaning transitioning from source to target language. 

Lyrics translation is a complex task which strives for inter-cultural comprehension of what makes a song suitable for singing. Prose translation, also called direct translation \cite{AttentionIsAllYouNeed,mbart,nllb}, differs greatly in its application from poetry and lyrics translation, as prose translation does not respect rules regarding rhythm and rhyme \cite{singability,PoetryTranslation}. A few attempts at automatic lyrics translation have been made by transforming standard music notation from one language to another, which shows promising results \cite{TonalLanguages,beauty-in-songs}. However, the necessity for standard music notation becomes a glaring restriction. From the perspective of a songwriter with interest in writing foreign-language lyrics, the creation of standard music notation is a labor-intensive task begging for automation. Therefore, to overcome the present limitations in adapting S2ST methods to the singing domain, we propose PolySinger: the first system for singing-voice to singing-voice translation (SV2SVT). PolySinger is a concatenated system of music information retrieval (MIR) technologies with the goal of directly translating a vocal performance in a source language into a synthetic vocal performance in a target language. PolySinger is made publicly available\footnote{\url{https://silasantonisen.github.io/polysinger/}}.

Automatic recognition of note-level events in a vocal melody is a complex and vaguely defined task \cite{vocano}. Nonetheless, standard music notation is required for lyrics translation, and as such, this work proposes a simple yet effective approach to defining note-level events by assistance from syllable alignment.

State-of-the-art (SOTA) in the following technologies are structured into a complete SV2SVT system for PolySinger: \emph{1)} automatic lyrics transcription, \emph{2)} phoneme-level lyrics alignment, \emph{3)} frame-level vocal melody extraction, \emph{4)} automatic lyrics translation, and \emph{5)} singing-voice synthesis. PolySinger is proposed as a concatenated solution instead of E2E to represent a modular framework facilitating research in SV2SVT. In this paper, PolySinger is presented for English to Japanese SV2SVT, which, to the best of our knowledge, also constitutes the first attempt at automatic lyrics translation from English to Japanese.

%While Japanese has moraic syllabaries in the form of kana where every character has a certain pronunciation, Japanese also uses kanji as logograms, that is, characters used to convey a certain meaning (see figure \ref{tab:kanji} for an explanation on the Japanese writing system). Japanese kanji have multiple readings depending on the context, and, as such, it becomes a challenging task to decide the pronunciation of a Japanese sentence. Furthermore, Japanese word segmentation and phrase break prediction are common problems in Japanese speech synthesis \cite{wordSegmentation,phraseBreak}. As such, solutions for circumventing challenges with the Japanese language in SV2SVT are also proposed in this work.

A series of native Japanese speakers participated in a mean opinion score (MOS) test to evaluate the perceptual quality of PolySinger for English to Japanese SV2SVT. Results show a promising fundamental structure for SV2SVT, but also that our translated Japanese lyrics have not yet reached ideal naturalness.

\section{Related Work}\label{sec:related work}
Convolutional neural networks and expanded pronunciation dictionaries have been used for automatic lyrics transcription in monophonic recordings \cite{monoALT}, along with time-delay neural networks in polyphonic recordings \cite{MSTRnet}. The latest reported SOTA in automatic lyrics transcription was achieved by adapting a Wav2Vec 2.0 \cite{wav2vec2} speech recognizer to the singing domain by transfer learning \cite{wav2vecALT}. However, in our preliminary tests we found the current SOTA speech recognition system, Whisper \cite{whisper}, to outperform the SOTA in automatic lyrics transcription \cite{wav2vecALT} when transcribing a vocal performance. Therefore, Whisper \cite{whisper} is used for automatic lyrics transcription in PolySinger.

Limited success has been achieved using speech alignment systems for lyrics alignment \cite{bootstraping}. However, great results have been obtained in word-level lyrics alignment by training a polyphonic acoustic model in \cite{DoesBackgroundMusicHelp}, but it is not until in \cite{PhonemeAlignment} that a direct attempt is made at phoneme-level lyrics alignment without sacrificing competitive performance in word-level alignment. Recent approaches have exploited the correlation between phoneme onset and note pitch by joint representation learning \cite{JointLearning} or cross-modal embedding in the audio and text domain through contrastive learning \cite{contrastive}. For PolySinger we use \cite{PhonemeAlignment} due to its documented performance in the specific task of phoneme-level lyrics alignment and accessibility to a pre-trained model.

Defining note-level events in a vocal melody is a complex task, and thus there is a lack of datasets and trained neural networks for note-level vocal melody extraction (VME) \cite{vocano}. On the other hand, frame-level VME is an extensively researched field with robust frameworks \cite{bittner2015melody,kum2016melody,2018VME}. Considering the high accuracy of most modern frame-level VME systems, \cite{2018VME} is used in PolySinger due to the streamlined implementation available through the MIR toolkit Omnizart \cite{omnizart}. Instead of defining the note-level events by VME, we define them by syllable-wise boundaries delimited by the phoneme-level lyrics alignments, and guide the pitch of those notes with frame-level VME.

In \cite{TonalLanguages}, a rule-based approach is suggested for translating from English to Chinese lyrics with respect to the original lyrics, melody and rhythm, as well as the tonal properties of Chinese. In \cite{beauty-in-songs}, a system for bidirectional translation between English and Chinese is proposed which incorporates an alignment decoder for determining the amount of syllables to write in the translation and how they should align to the melody. Additionally in \cite{beauty-in-songs}, the evaluation process of the system is assisted by synthesizing the translation via SVS.
For PolySinger, we take inspiration from \cite{TonalLanguages} by going for a simple rule-based approach for English to Japanese lyrics translation due to data scarcity and an interest in unraveling the implications of processing Japanese lyrics. To do so, we exploit the pre-trained SOTA model for multi-lingual translation \texttt{nllb-200} \cite{nllb} by transferring it to the singing domain.

Early work on SVS created concatenated singing libraries of sampled vocal sounds in a wide range of pitches, from which a synthesizer chose the samples for synthesis based on a musical score \cite{concatenation,vocaloid}. More recent approaches use acoustic models trained on vocal performances from a singer to replicate the way he/she would perform a song given a musical score \cite{liu2022diffsinger}. Furthermore, cross-lingual synthesis has become possible even when only training on mono-lingual singers \cite{crosssinger}. The open-source scene has entered the SVS consumer market, e.g., by use of the ENUNU\footnote{\url{https://github.com/oatsu-gh/ENUNU}} plugin to enable usage of the NNSVS toolkit \cite{nnsvs} in the OpenUTAU editor. Synthesizer V\footnote{\url{https://dreamtonics.com/synthesizerv/}} is gathering a common consensus of being one of the best consumer products for SVS with a wide range of high-quality neural singing libraries capable of cross-lingual synthesis in a user-friendly environment with scripting capabilities. Therefore, Synthesizer V is used for SVS in PolySinger. Similarly to \cite{beauty-in-songs}, we synthesize the translated lyrics, but we want to emphasize that, \emph{differently from \cite{beauty-in-songs}, PolySinger automates the intermediate link between automatic lyrics translation and SVS}.

\section{Proposed Singing-Voice to Singing-Voice Translation System}\label{sec:Singing-voice to Singing-Voice-Synthetis System}

\begin{figure*}[t]
 \centerline{\framebox{
 \includegraphics[width=\textwidth]{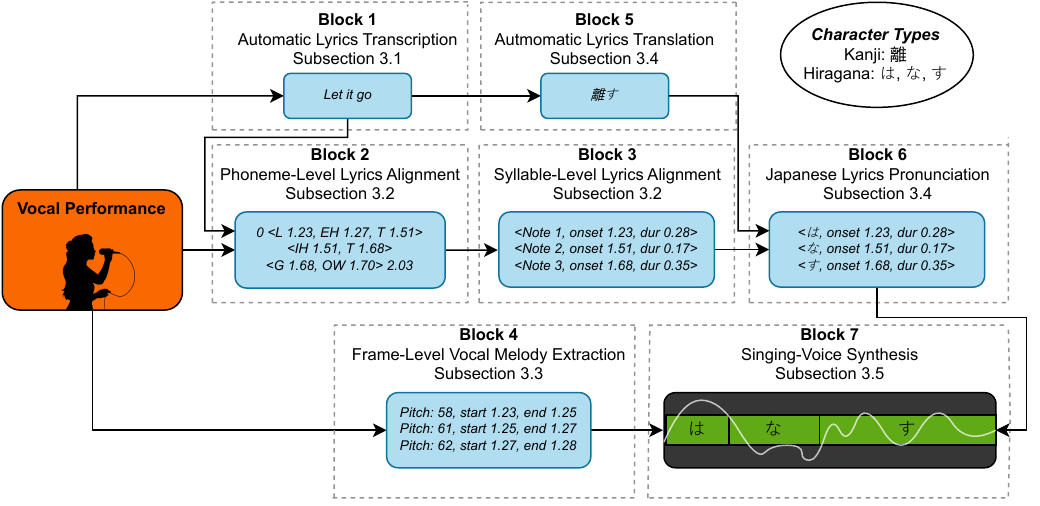}}}
 \caption{Overview of our proposed SV2SVT system, PolySinger. Provided an English vocal performance, a synthetic vocal performance is created in Block 7, defined by notes with onsets, durations and Japanese lyrics, guided by a frame-level melody. Every numeric value is in seconds and ``< >'' illustrates the boundaries of notes. The process of segmenting words into syllables is illustrated in Table \ref{tab:syllables}. Fundamentals of Japanese writing are explained in Subsection \ref{subsec:translation}, and the process of converting kanji to hiragana is illustrated in Table \ref{tab:kanji}.}
 \label{fig:pipeline}
\end{figure*}

%Overview of our proposed SV2SVT system, PolySinger. An initial English vocal performance is sent into Block 1 for automatic lyrics transcription which gets aligned to the vocal performance at a phoneme-level in Block 2. The phonemes are concatenated into syllables with respect to the CMUdict (this process is further demonstrated in Table \ref{tab:syllables}). We define notes as syllables, thus the temporal alignments of phonemes belonging to a syllable determine the onset and duration of the note in Block 3. The melody to go along with the notes is extracted at a frame-level in Block 4 from the vocal performance. The lyrics provided by Block 1 are translated into Japanese in Block 5. Multiple translations are generated, and the one that suits the note sequence the best is picked. Japanese kanji characters are decoded into hiragana (Japanese characters and the conversion between them is explained in Subsection \ref{subsec:pronunciation}). Each hiragana get assigned to a note in Block 6 and provded to the SVS system in Block 7 along with the frame-level melody

Figure \ref{fig:pipeline} illustrates a flowchart of the proposed SV2SVT system. This section will systematically break down the technology, implementation and functionality of each block presented in this figure.

\subsection{Automatic Lyrics Transcription}\label{subsec:lyrics transcription}
Whisper \cite{whisper} is a Transformer-based model \cite{AttentionIsAllYouNeed} originally pre-trained on 680k hours of weakly-labeled audio for multi-task learning; there among the main task being multi-lingual automatic speech recognition. The most recent checkpoint, \texttt{Whisper-large-V3}, is trained on 1M hours of weakly-labeled audio and 4M hours of audio which was pseudo-labeled by \texttt{Whisper-large-V2}. We collect \texttt{Whisper-large-V3} from HuggingFace\footnote{\url{https://huggingface.co/openai/whisper-large-v3}} for automatic lyrics transcription. The model has 1,550M parameters and was trained for 2 epochs on the dataset. We have not fine-tuned Whisper on singing data due to Whisper's great ability to generalize across several domains. To keep the memory usage of Whisper within $\sim$8 GB, a chunking algorithm segments the vocal performance into 30-second segments which are processed individually with a batch size of 4. Block 1 in Figure \ref{fig:pipeline} is facilitated by Whisper to transcribe an English string of text from an English vocal performance.

\subsection{Phoneme-Level Lyrics Alignment and Syllable-Level Lyrics Alignment}\label{subsec:alignment}
In Western languages, poetry and lyrics are very reminiscent of each other. Poetry has a rhythmic structure called meter. This structure can be dissected into a syllabic pattern \cite{greene2010automaticanalysis}. Therefore, in this work, we define the onset and duration of notes by aligning the sung syllables to the vocal performance. To obtain syllable-level lyrics alignments, we first align the sequence of phonemes present in the vocal performance. The phoneme sequence is extracted with the pre-trained phoneme-level lyrics aligner informed in \cite{PhonemeAlignment}. This model is a deep neural network trained for joint phoneme-level lyrics alignment and singing-voice separation. Text and audio are encoded separately. Text features and audio features are aligned by dynamic time warping-attention to minimize the total distance between audio frames and phonemes. The list of possible phonemes is provided by CMUdict\footnote{\url{http://www.speech.cs.cmu.edu/cgi-bin/cmudict}}. Block 2 in Figure \ref{fig:pipeline} uses the phoneme-level lyrics aligner to align the transcribed lyrics provided by Block 1 to the vocal performance. The phoneme alignments are informed in seconds. The phoneme sequence is concatenated into syllables by simple rules: \emph{1)} it is assumed that each phoneme corresponding to a vowel makes an individual syllable, and \emph{2)} consonants are merged with their closest neighboring vowel, gravitating towards the rightmost vowel in case of both neighboring phonemes corresponding to vowels. The process of breaking a word into phonemes according to CMUdict and concatenating them into syllables is illustrated in Table \ref{tab:syllables}. In Block 3 of Figure \ref{fig:pipeline}, in order to perform syllable-level lyrics alignment, we define the onset of a note as the start of the first phoneme in a syllable, and we define the duration as the time difference between the onset and the end of the last phoneme in the syllable.

\begin{table}
  \centering
  {\begin{tabular}{l|c}
    \toprule[1pt]\midrule[0.3pt]
    \emph{Word} & BLUEBERRY \\ \midrule
    \emph{Phonemes} & B \hspace{0.2cm} L \hspace{0.2cm} UW1 \hspace{0.2cm} B \hspace{0.2cm} EH2 \hspace{0.2cm} R \hspace{0.2cm} IY0 \\ \midrule
    \emph{Syllables} & BLUW \hspace{0.2cm} BEH \hspace{0.2cm} RIY\\
    \midrule[0.3pt]\bottomrule[1pt]
  \end{tabular}}
      \caption{Example of the word ``blueberry'' being deconstructed into phonemes with respect to CMUdict and reconstructed into syllables. An integer ranging 0-2 is associated with each vowel to indicate the type of vowel stress.}
  \label{tab:syllables}
\end{table}

\subsection{Vocal Melody Extraction}\label{subsec:VME}

The notes in Block 3 of Figure \ref{fig:pipeline} with timings defined by phoneme boundaries are not individually associated with a unique pitch. Instead, to preserve the melody from the vocal performance as much as possible, the melody is extracted at a frame level. This frame-level pitch contour is used to automate the pitch over time for the note sequence. The notes are set to a standard pitch of 60, and the contour is used to describe the deviation from 60 at each frame. The frame-level VME system presented in \cite{2018VME} deploys a deep convolutional neural network for semantic segmentation across a time-frequency image. Additionally, a progressive neural network is used for cross-domain transfer learning between the audio domain (frequencies) and the symbolic domain (pitch). Block 4 in Figure \ref{fig:pipeline} utilizes this model for extracting the frame-level melody contour as pitches from the vocal performance. The start and end of pitches are in seconds. The model is used through the Omnizart toolkit \cite{omnizart}.

\subsection{Automatic Lyrics Translation and Japanese Lyrics Pronunciation}\label{subsec:translation}
\texttt{nllb-200} \cite{nllb} is a Transformer \cite{AttentionIsAllYouNeed}-based mixture-of-experts (MoE) multi-lingual translation model that achieves SOTA results in many languages. This success is largely owed to the parallel development of datasets: \emph{1)} the expertly-annotated Flores-200 dataset (which consists of 3,001 English sentences translated into 204 languages), and \emph{2)} automatically-generated datasets by web-scraping for either mono-lingual sentences with high probability of being each other's translation, or mono-lingual sentences for back-translation. The biggest model, \texttt{nllb-200-MoE}, has 54B parameters, which is infeasible to run locally. Therefore, we collect the smaller checkpoint \texttt{nllb-200-distilled-600M} from HuggingFace\footnote{\url{https://huggingface.co/facebook/nllb-200-distilled-600M}} and fine-tune the model for English to Japanese lyrics translation. Since neither high-quality nor high-quantity dataset of paired English and Japanese lyrics exists, we take inspiration from \cite{TonalLanguages} and scrape the web for lyrics translations that are not necessarily singable. Our dataset consists of $\sim$213k paired lines, thereof $\sim$80\% ($\sim$20\%) being Japanese$\Rightarrow$English (English$\Rightarrow$Japanese)\footnote{Both English$\Rightarrow$Japanese and Japanese$\Rightarrow$English are collected from \url{https://lyricstranslate.com/}. Extra Japanese$\Rightarrow$English is also collected from \url{https://www.animelyrics.com/}. More information about this dataset can be found at \url{https://silasantonisen.github.io/polysinger/}}. With PolySinger, we perform English$\Rightarrow$Japanese translation, so we invert the Japanese$\Rightarrow$English lyrics pairs into English$\Rightarrow$Japanese lyrics pairs. We hypothesize that this inversion does not raise an issue, but might in fact incentivize the model to produce high-quality Japanese lyrics even when provided with low-quality English lyrics. Besides, unlike in \cite{TonalLanguages}, we attempt fine-tuning with no prior self-supervised training on mono-lingual lyrics due to our larger paired dataset.

Ideally, the output of our fine-tuned model should have the same amount of syllables as established in Block 3 of Figure \ref{fig:pipeline}. However, counting syllables is not as simple in Japanese as in English. Japanese has moraic syllabaries in the form of kana. Kana characters have specific pronunciations that take up one mora (Japanese syllable) each. Japanese also uses kanji as logograms, that is, characters that convey a certain meaning. Kanji characters have multiple readings depending on the context, and, as such, it becomes a challenging task to decide the pronunciation of a Japanese sentence. To get the correct pronunciation of kanji characters, \texttt{pyKAKASI}\footnote{\url{https://codeberg.org/miurahr/pykakasi}} is used to decode kanji into their hiragana (a type of kana) readings. An illustration of the relation between kanji and Japanese pronunciation can be seen in Table \ref{tab:kanji}. \texttt{pyKAKASI} is dictionary-based, and it can therefore be difficult to convert sentence-wise instead of word-wise. Japanese does not use blank space to separate words, therefore, we use \texttt{Nagisa}\footnote{\url{https://github.com/taishi-i/nagisa}}, a recurrent neural network trained for Japanese word segmentation.

During inference, a beam search is applied to the output of our fine-tuned \texttt{nllb-200-distilled-600M} with as many beams as memory will allow ($\sim$50 beams in our tests). The beams are biased towards a token count lower than the number of syllables in Block 3 of Figure \ref{fig:pipeline} due to a kanji always corresponding to at least one syllable. Each generated sentence becomes word-separated with \texttt{Nagisa} and the kanji are converted into hiragana readings with \texttt{pyKAKASI}. The sentence with the lowest non-negative difference between mora count and syllable count gets selected and assigned to the notes in Block 6 of Figure \ref{fig:pipeline}.

\begin{table}
  \centering
    \begin{tabular}{l|c}
    \toprule[1pt]\midrule[0.3pt]
    \emph{Kanji character} & \begin{CJK}{UTF8}{min} 離 \end{CJK} \hspace{0.3cm} \\ \midrule
    \emph{Hiragana readings} & \begin{CJK}{UTF8}{min} り \end{CJK} \hspace{0.3cm} \begin{CJK}{UTF8}{min} はな \end{CJK} \\ \midrule
    \emph{Roman readings} & \hspace{0.2cm} RI \hspace{0.3cm} HA NA\\ \midrule
    \emph{Mora count} & \hspace{-0.25cm} 1 \hspace{0.8cm} 2 \\
    \midrule[0.3pt]\bottomrule[1pt]
  \end{tabular}
      \caption{Example of two possible hiragana readings for a kanji character.}
  \label{tab:kanji}
\end{table}

\subsection{Singing-Voice Synthesis}\label{subsec:svs}
Synthesizer V is a SVS system with growing popularity among musicians. The technology behind Synthesizer V is kept proprietary. Based on related literature \cite{liu2022diffsinger,nnsvs}, it is assumed that AI singing-voices in Synthesizer V are acoustic models trained on phoneme-level annotated vocal performances. With this training scheme, the model recognizes patterns in a singer's vocal performances, e.g., articulation of phoneme sequences, transitions between pitches and tendencies to use vibrato. Additionally, Synthesizer V AI voices have parameters for vocal modes, which can be included in training by annotating vocal samples with a singing style, e.g., nasal, powerful, soft, and whisper. AI singing-voices are usually only trained on vocal performances by a mono-lingual or bilingual singer, however AI voices in Synthesizer V are capable of cross-lingual synthesis in English, Japanese, Mandarin Chinese, Cantonese, and, recently, Spanish. It is assumed, based on related literature \cite{crosssinger}, that cross-lingual synthesis is achieved by unifying phoneme representations across languages with the international phonetic alphabet and training on data labeled with language identification such that the acoustic model can learn language-specific features. As illustrated in Figure \ref{fig:pipeline}, the notes with Japanese lyrics provided by Block 6 are plotted into Synthesizer V at a standard pitch of 60. The vocal contour provided by Block 4 is used to automate the deviation from pitch 60 over time. We use the AI singing-voice Mai in Synthesizer V to generate the Japanese vocal performance.

\section{Experiments}\label{sec:experiments}
Objective measures in machine translation such as BLEU \cite{papineni2002bleu} are typically used for word-wise similarity with respect to a ground truth. Such a method does not suit lyrics translation as there should rather be a focus on semantic interpretation rather than precise word choice. Moreover, PolySinger has to be evaluated on the overall performance achieved for English$\Rightarrow$Japanese SV2SVT rather than solely on the translation quality. Therefore, we evaluate PolySinger subjectively by means of a MOS test.

\begin{table}
  \centering
  \resizebox{\linewidth}{!}{\begin{tabular}{l|cccccc}
    \toprule[1pt]\midrule[0.3pt]
    \emph{Score} & 1 & 2 & 3 & 4 & 5 \\ \midrule
    \emph{Meaning} & Very poor & Poor & Neutral & Good & Very good \\
    \midrule[0.3pt]\bottomrule[1pt]
  \end{tabular}}
      \caption{Five-point scale for MOS test.}
  \label{tab:scores}
\end{table}

\begin{table}
  \centering
  \resizebox{\linewidth}{!}{\begin{tabular}{c|l}
    \toprule[1pt]\midrule[0.3pt]
    \textbf{ID} & \textbf{Question} \\ \midrule
    Q1 & How much sense do the lyrics make? \\
    Q2 & How natural is the Japanese used in the lyrics? \\
    Q3 & How well is the meaning of the original lyrics preserved? \\
    Q4 & How singable are the generated lyrics? \\
    Q5 & How well are the lyrics and melody aligned? \\
    Q6 & What is the overall quality of the generated Japanese singing? \\
    \midrule[0.3pt]\bottomrule[1pt]
  \end{tabular}}
  \caption{Questions asked to the test subjects in the MOS test of Section \ref{sec:experiments}.}
  \label{tab:questions}
\end{table}

\begin{table*}
  \centering
  \renewcommand{\arraystretch}{0.8}
  \begin{tabular}{l|cccccc}
    \toprule[1pt]\midrule[0.3pt]
    \textbf{System / Question} & \textbf{Q1} & \textbf{Q2} & \textbf{Q3} & \textbf{Q4} & \textbf{Q5} & \textbf{Q6} \\ \midrule
    Baseline & 2.53 $\pm$ 0.49 & 2.57 $\pm$ 0.48 & 2.47 $\pm$ 0.44 & 2.40 $\pm$ 0.41 & 2.50 $\pm$ 0.52 & 2.33 $\pm$ 0.45 \\
    Fine-tuned & 2.17 $\pm$ 0.46 & 2.30 $\pm$ 0.48 & 2.10 $\pm$ 0.44 & 2.23 $\pm$ 0.44 & 2.10 $\pm$ 0.40 & 2.13 $\pm$ 0.41 \\
    \midrule[0.3pt]\bottomrule[1pt]
  \end{tabular}
  \caption{MOS quality test results, broken down by question, with 95\% confidence intervals.}
  \label{tab:results}
\end{table*}

\begin{table}
  \centering
  \resizebox{\linewidth}{!}{\begin{tabular}{l|cccccc}
    \toprule[1pt]\midrule[0.3pt]
    \emph{Question} & Q1 & Q2 & Q3 & Q4 & Q5 & Q6 \\ \midrule
    \emph{$p$-value} & 0.228 & 0.389 & 0.115 & 0.509 & 0.279 & 0.557 \\
    \midrule[0.3pt]\bottomrule[1pt]
  \end{tabular}}
  \caption{$p$-values, broken down by question, from a Wilcoxon rank-sum test comparing MOS scores from Baseline and Fine-tuned.}
  \label{tab:p-values}
\end{table}

\begin{table}
  \centering
  \resizebox{\linewidth}{!}{\begin{tabular}{c|l}
    \toprule[1pt]\midrule[0.3pt]
    \textbf{ID} & \textbf{Comment} \\ \midrule
    C1 & Incorrect readings of kanji \\
    C2 & Usage of keigo in casual language \\
    C3 & Direct translations where interpretations are needed \\
    C4 & Occasionally, lyrics are not entirely translated \\
    C5 & Both intra- and inter-word separation at unnatural places \\
    C6 & Missing keywords important to the song \\
    C7 & Wrong word order \\
    C8 & Improper mixture of feminine and masculine language \\
    \midrule[0.3pt]\bottomrule[1pt]
  \end{tabular}}
  \caption{Comments from discussions with the test subjects on essential improvements that could lead to more natural synthetic Japanese singing.}
  \label{tab:comments}
\end{table}

\subsection{Methodology}\label{subsec:methodology}
Six native Japanese speakers participated in a MOS test to evaluate the perceptual quality of English$\Rightarrow$Japanese SV2SVT using PolySinger on 5 different vocal performances. All test subjects were females ranging from 24 to 39 years old with no hearing impairment. The test subjects were asked to self-report their English speaking level. Two participants reported complete fluency (5/5), one reported near fluency (4/5), two more indicated advanced comprehension (3/5), and the final one reported intermediate comprehension (2/5). Using the inference procedure described in Subsection \ref{subsec:translation}, PolySinger was alternately tested with the original \texttt{nllb-200-distilled-600M} (\emph{Baseline}) and our fine-tuned \texttt{nllb-200-distilled-600M} (\emph{Fine-tuned}) on every vocal performance. The test subjects were asked to first listen to an English vocal performance, followed by the synthetic performances generated by the two PolySinger versions (i.e., Baseline and Fine-tuned). Participants were not informed which synthetic vocal performance was generated by which system variant. Using the 5-point scale shown in Table \ref{tab:scores}, the test subjects were asked to assess each generated performance by the 6 MOS questions displayed in Table \ref{tab:questions}. The average time a participant spent on the evaluation was 53 min. The audio samples used for evaluation can be accessed here\footnote{\url{https://silasantonisen.github.io/polysinger/}}.

After the participants submitted their MOS scores, we additionally had a brief discussion with them individually about their general opinions and observations.

\begin{figure*}
    \centering
    \includegraphics[width=0.48\linewidth]{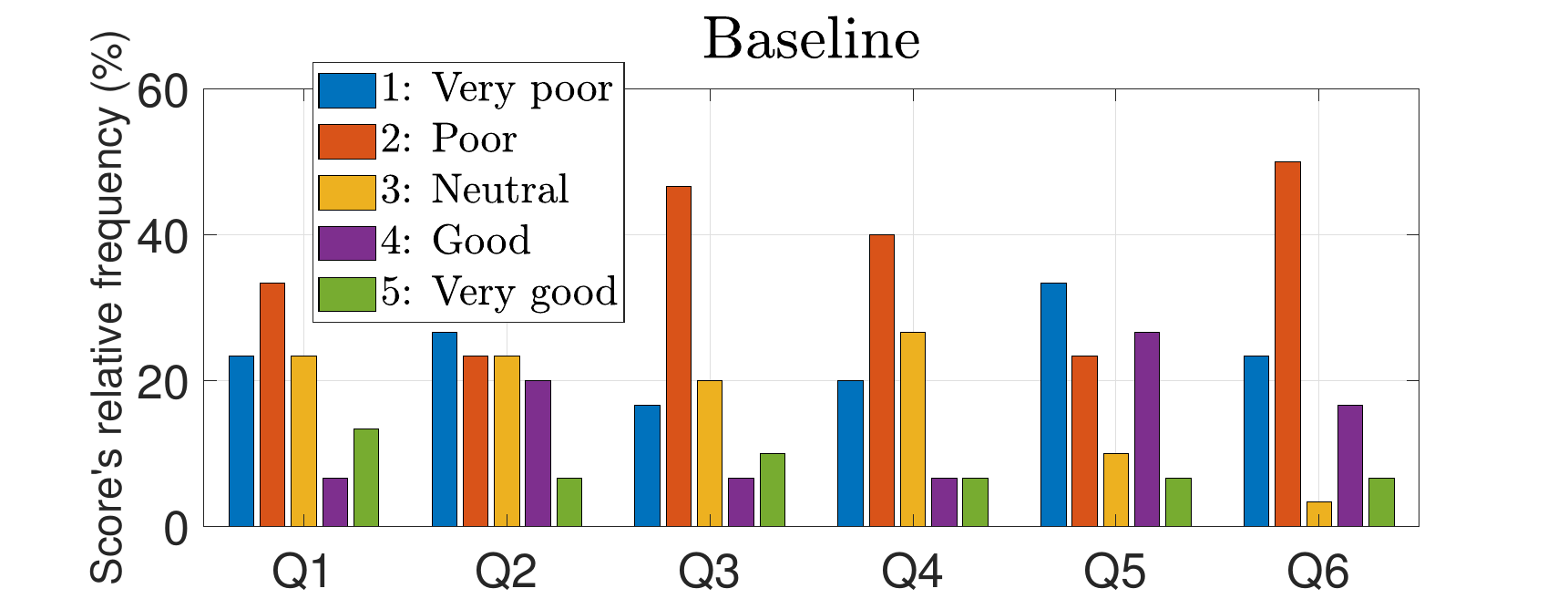} \hfill \includegraphics[width=0.48\linewidth]{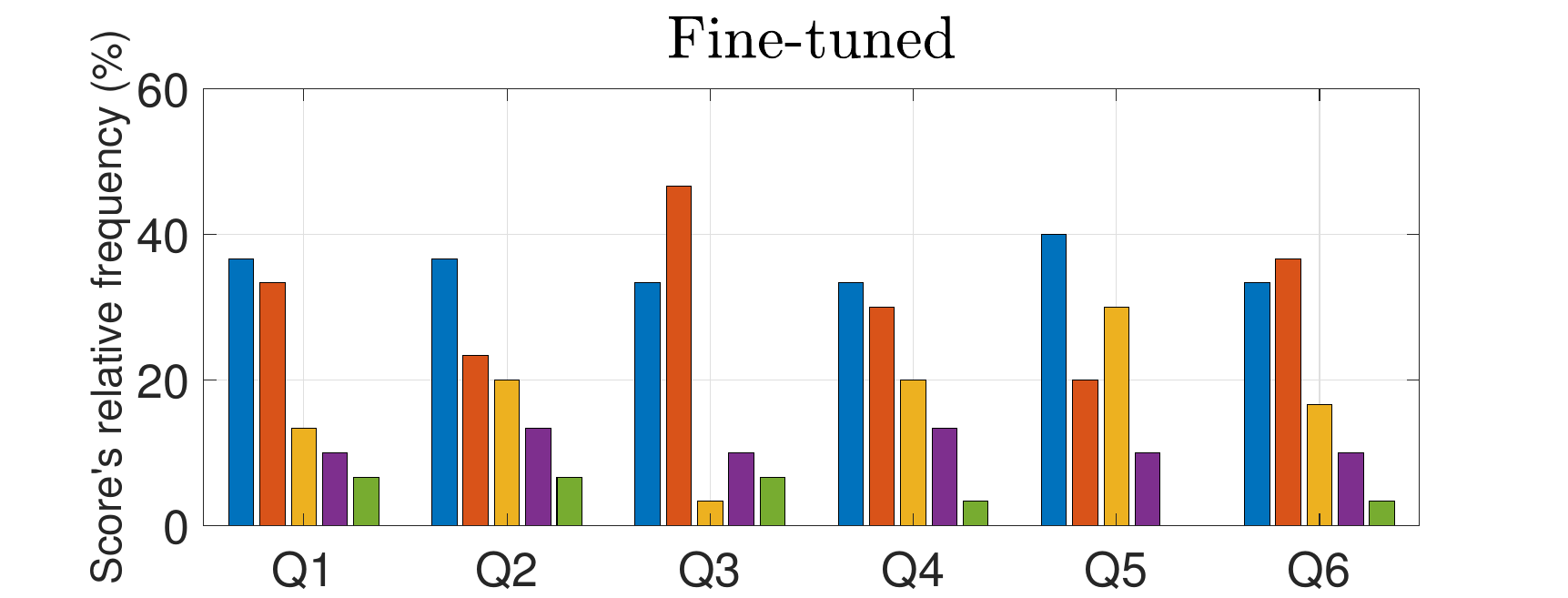}
    \caption{Bar plots representing per-question score's relative frequency from the MOS quality test for Baseline (left) and Fine-tuned (right).}
    \label{fig:bar_plots}
\end{figure*}

\subsection{Results}\label{subsec:results}
Table \ref{tab:results} shows the MOS test results along with 95\% confidence intervals from the Student’s $t$-distribution \cite{Blachman}. Both system variants (i.e., Baseline and Fine-tuned) lie somewhere between poor and neutral in all 6 MOS questions Q1--Q6. The relatively large confidence intervals in Table \ref{tab:results} suggest a high variance in opinion scores. We investigate this variance in Figure \ref{fig:bar_plots} by representing per-question score's relative frequency. While it is true that the majority of opinion scores lies in the mid-to-low end of the spectrum, several evaluations have also resulted in good or very good opinion scores. This emphasizes the very subjective nature of the SV2SVT problem.

Given a MOS question Q1--Q6, we determine if there is a statistically significant difference between the opinion scores for Baseline and Fine-tuned. A Kolmogorov-Smirnov test \cite{massey1951kolmogorov} generally rejects, at a standard significance level of 5\%, the null hypothesis that our opinion score sample populations follow Gaussian distributions. Therefore, we use a Wilcoxon rank-sum test \cite{wilcoxon1945individual} to determine whether there are statistically significant differences in MOS between the two system variants. The $p$-values shown in Table \ref{tab:p-values} demonstrate that the performance of the two systems is rather equivalent. Specifically, these $p$-values indicate that there are no statistically significant differences between Baseline and Fine-tuned at a standard significance level of 5\% given any of the 6 MOS questions.

During discussions conducted after the MOS test, the test subjects generally conveyed a positive reaction towards SV2SVT being possible with PolySinger. However, as anticipated, the participants mainly assessed PolySinger by the naturalness of the Japanese language used in the context of singing and the pronunciation of words. The most recurring observations from the participants, that were suggested as crucial improvements needed for the pursuit of natural Japanese singing, are summarized in Table \ref{tab:comments}. In the next section, we will discuss the comments in Table \ref{tab:comments} as to why our processing and synthesis of Japanese might not have been of ideal quality, along with our plan for improving them in future. Moreover, the statistically insignificant difference between Baseline and Fine-tuned is also discussed along with techniques and technologies that may assist in improving PolySinger.

\section{Discussion}\label{sec:discussion}

%The purpose of this paper has mainly been to unravel the possibilities of adapting conventional S2ST to the singing domain. To answer the question at hand, SOTA methods in music information retrieval have been cascaded into a complete system for SV2SVT. An alternative approach would be to develop an end-to-end solution similar to that of recent approaches in S2ST \cite{textlessS2ST, seamlessm4t2023}. However, several complications will arise when considering an end-to-end SV2SVT system: \emph{1)} It is difficult to identify a bottleneck in an end-to-end system relative to a cascaded system, and an end-to-end solution also offers less control. This is a problem given that the literature has not yet questioned what the necessities for a high quality SV2SVT system are, which is one of the major goal of this paper. \emph{2)} End-to-end S2ST technology is still in its infancy, with a scarce amount of literature and open-source software. Developing an end-to-end SV2SVT system would either require the construction of a new complex structure, performing alteration of an existing model, or transfer learning from an existing model to the singing domain. \emph{3)} Paired English-Japanese text-based lyrics are scarce, but so is it for pairs of audio-based data, thus not alleviating the data scarcity problem, but introducing new complications caused by a different domain with less surrounding literature.

To produce natural Japanese speech synthesis, the front-end of a text-to-speech system requires phonetic and prosodic features \cite{fujimoto2019impacts}. Phonetic features, i.e., pronunciations, are typically acquired by grapheme-to-phoneme (G2P) conversion, and prosodic features, i.e., rhythm and intonation, are in Japanese typically acquired by phrase break prediction and accent estimation \cite{kakegawa2021phonetic, phraseBreak, kurihara2021prosodic}. G2P conversion is particularly difficult in Japanese, since kanji characters can have multiple pronunciations. As indicated by our test subjects (C1 in Table \ref{tab:comments}) and discussed in \cite{kakegawa2021phonetic}, the accuracy obtained by dictionary-based G2P conversion in Japanese is not satisfactory. Japanese has no word separators, which also makes it difficult to determine phrase breaks. In our work, we performed word segmentation with \texttt{Nagisa} to avoid intra-word breaking, and attempted to define phrase breaks as the pauses transcribed by lyrics alignment on an English vocal performance. However, according to our test subjects (C5 in Table \ref{tab:comments}), these methods yielded limited success. As future work, we will investigate the adaptation of SOTA methodologies in Japanese text-to-speech to SV2SVT such as phrase break prediction with large language models (LLMs) \cite{phraseBreak} and G2P conversion via machine translation \cite{kakegawa2021phonetic}.

In \cite{TonalLanguages}, they demonstrated an improvement in automatic lyrics translation by fine-tuning on paired lyrics that were not necessarily singable, but also by pre-training on mono-lingual lyrics. In this work, we avoided pre-training on mono-lingual lyrics and only fine-tuned on paired lyrics that were not necessarily singable, which resulted in no statistically significant improvement with respect to the baseline model (see Table \ref{tab:p-values}). We applied a beam search to find translated lyrics that fit well into the syllable count of the original lyrics. The selected lyrics were occasionally not a full translation of the original lyrics (C4 in Table \ref{tab:comments}). Apart from the use of keigo (honorific language) being inappropriate for the inherent casual nature of song lyrics (C2 in Table \ref{tab:comments}), we conjecture that keigo could also be a major cause of incomplete lyrics translations. This is because keigo will usually incorporate more characters than casual language, which means that it will be harder to fit the lyrics into the fixed syllable count.

In \cite{beauty-in-songs}, they achieve SOTA results by training on a dataset created by back-translating mono-lingual lyrics and automatically aligning automatically-generated melodies that fit both the source and target lyrics. As future work, creating such a dataset and training an alignment decoder similarly to \cite{beauty-in-songs} could very well be adapted to Japanese. However, we hypothesize that translation systems have an inherent limitation towards cross-lingual songwriting that hinders them from rivaling professional human translators due to a lack of abstract interpretation and ``imagination''. Hence, as future work, we will also investigate the usage of LLMs for sentiment analysis and feature extraction to exploit poetry/lyrics generation models. By lyrics generation, guided by keyword spotting, we can also address the issue of missing keywords (C6 in Table \ref{tab:comments}).

\section{Conclusion}\label{sec:conclusion}
The goal of this paper has been to adapt conventional S2ST to the singing domain. To do so, we have built the first SV2SVT system, PolySinger, by cascading SOTA MIR technologies facilitating a modular tool for extended research in SV2SVT. We have conducted a MOS test with native Japanese speakers to evaluate PolySinger's performance for English to Japanese SV2SVT. Results indicate that we have created a fundamentally-coherent structure for SV2SVT, but the translation of English lyrics into Japanese and the automatic synthesis of it is not yet natural enough. To further develop SV2SVT, our future work will investigate ---to facilitate creative lyrics generation--- the usage of sentiment analysis and feature extraction for abstract meaning representation of lyrics as opposed to translation. Finally, we will also investigate the necessities for autonomous generation of natural Japanese lyrics.

\section{Acknowledgments}
This work has been funded by the Spanish Ministry of Science and Innovation under the ``Ram\'on y Cajal'' programme (RYC2022-036755-I). In addition, we want to express our heartfelt appreciation to the test subjects who voluntarily contributed to this study by participating in the perceptual quality test.

% For bibtex users:
\bibliography{ISMIRtemplate}

% For non bibtex users:
%\begin{thebibliography}{citations}
% \bibitem{Author:17}
% E.~Author and B.~Authour, ``The title of the conference paper,'' in {\em Proc.
% of the Int. Society for Music Information Retrieval Conf.}, (Suzhou, China),
% pp.~111--117, 2017.
%
% \bibitem{Someone:10}
% A.~Someone, B.~Someone, and C.~Someone, ``The title of the journal paper,''
%  {\em Journal of New Music Research}, vol.~A, pp.~111--222, September 2010.
%
% \bibitem{Person:20}
% O.~Person, {\em Title of the Book}.
% \newblock Montr\'{e}al, Canada: McGill-Queen's University Press, 2021.
%
% \bibitem{Person:09}
% F.~Person and S.~Person, ``Title of a chapter this book,'' in {\em A Book
% Containing Delightful Chapters} (A.~G. Editor, ed.), pp.~58--102, Tokyo,
% Japan: The Publisher, 2009.
%
%
%\end{thebibliography}

\end{document}